# Can simple transmission chains foster collective intelligence in binary-choice tasks?

**Mehdi Moussaïd[1*] and Kyanoush Seyed Yahosseini[1]**

[1] Center for Adaptive Rationality, Max Planck Institute for Human Development, Berlin, Germany.
*Corresponding author: moussaid@mpib-berlin.mpg.de

# Abstract

In many social systems, groups of individuals can find remarkably efficient solutions to complex cognitive problems, sometimes even outperforming a single expert. The success of the group, however, crucially depends on how the judgments of the group members are aggregated to produce the collective answer. A large variety of such aggregation methods have been described in the literature, such as averaging the independent judgments, relying on the majority or setting up a group discussion. In the present work, we introduce a novel approach for aggregating judgments — the transmission chain — which has not yet been consistently evaluated in the context of collective intelligence. In a transmission chain, all group members have access to a unique collective solution and can improve it sequentially. Over repeated improvements, the collective solution that emerges reflects the judgments of every group members. We address the question of whether such a transmission chain can foster collective intelligence for binary-choice problems. In a series of numerical simulations, we explore the impact of various factors on the performance of the transmission chain, such as the group size, the model parameters, and the structure of the population. The performance of this method are compared to those of the majority rule and the confidence-weighted majority. Finally, we rely on two existing datasets of individuals performing a series of binary decisions to evaluate the expected performances of the three methods empirically. We find that the parameter space where the transmission chain has the best performance rarely appears in real datasets. We conclude that the transmission chain is best-suited for other types of problems, such as those that have cumulative properties.

# Introduction

Collective intelligence refers to the ability of groups of individuals to find solutions to complex problems. The term "collective intelligence" – also referred to as "swarm intelligence" or "collective problem-solving" – is used in a surprisingly large diversity of interdisciplinary domains. These include the collective behaviour of animal swarms [1–3], the processes underlying group discussions and brainstorming in business and industry [4,5], the wisdom-of-the-crowds and other methods for combining judgments [6–10], the design of artificial multi-agents systems in robotics and biomimetics [11,12], the behaviour of pedestrian crowds [13–15], networked experiments in social computing [16–18], citizen science [19,20], and numerous online collaborative projects such as Wikipedia and Threadless [21]. All these seemingly disparate domains share the same overarching principle: The collective solution that is produced by the group results from the *aggregation* of every individual's judgment. The central question is therefore: What is the best way to aggregate a multitude of individual solutions?

A very large number of such aggregation methods exist in the literature. These can be classified into three major families, depending on the nature of the interactions between the group members.

The first family of aggregation methods concerns situations where group members *do not interact* with each other. In this case, a third-party decision-maker collects the independent solutions of each individual and aggregates them to produce the collective solution [8,22,23]. Various aggregation methods have been proposed, such as a simple average, a majority rule, a quorum-based decision, or aggregation rules taking into account the confidence level of the individuals, such as the maximum-confidence slating or the confidence-weighted majority [8,23–25]. For instance, the wisdom-of-the-crowds is a prominent method that consists in computing the mean or the median value of all individuals' judgments [6]. More sophisticated aggregation methods can successfully combine individual solutions even for complex, multi-dimensional problems, such as the traveling salesman problem [26], or ranking problems [27]. In any case, all individual solutions must be independent from one another, since direct or indirect sources of social influence often undermine the power of the aggregation [7].

The above family of aggregation methods has been studied for more than a century [6]. More recently, the understanding of self-organised social systems has demonstrated that groups of people and swarms of animal are also capable to solve complex problems collectively *without* the intervention of a third-party decision-maker. In this case, the aggregation of the information is supported by the interactions among the individuals [13]. In the '90s, biologists studying animal swarms have classified these interactions in two types: direct and indirect interactions [11,12,28,29], which turned out to be applicable to human groups as well. Formally, direct interaction refers to situation where individuals collect information directly from other individuals. In contrast, indirect interaction describes situations where individuals collect information from the collective solution that is emerging, which indirectly reflects what other individuals have done. In the following paragraphs, we describe and illustrate these two processes.

*Direct interactions* constitute the second family of aggregation methods, and refer to groups in which the individuals collect information directly from other individuals, through visual, acoustic, or electronic signals. Over repeated interactions, the information that each group member possesses flows from individual to individual, and eventually gives rise to an adaptive collective response. In biological systems, fish schools and bird flocks are typical examples of systems that address complex problems through direct interactions. Animals perceive visual and acoustic cues from their neighbors and adapt their behavior accordingly. The propagation of information from individual to individual gives rise to collective patterns, for instance, when detecting and avoiding predators [30–32]. In humans, situations where people directly interact with each other are typically group discussions. During group discussions, all members can freely exchange ideas, suggestions and information, and try to come up with a joint solution to a given problem. Depending on the structure and the composition of the group, the discussion can constitutes a powerful mean to aggregate judgments [4,33–35], but this method is also subject to various undesirable effects such as opinion herding, groupthink, and the hidden profile effect [5,36,37]. Direct interactions can also take place in social networks. In networked experiments, for instance, all group members can observe the behavior and solution of their neighbours and adapt to it. Over successive rounds of interactions, the group often converges to a collective solution [16,17]. This method has been successfully applied to the graph-colouring problem [16], and other collective exploration problems [18,38].

Finally, the third family of aggregation methods contains situations where the group members *indirectly interact* with one another. That is, individuals do not directly collect information from each other, but from the current state of the collective solution. In these situations, individuals typically work simultaneously or sequentially on a common collective solution. At any moment, the current state of the collective solution drives the subsequent actions of the individuals. In biological systems, this type of interaction is best illustrated by the nest construction in social insects [39–41]: Each piece of construction material that an insect deposits on the collective construction motivates the other insects to add another piece to the structure. Each individual collects information from the current state of the structure, which reflects the cumulative actions of previous workers. Over time, this process gives rise to small heaps, which become columns and eventually form the complex structure of the nest. In human groups, Wikipedia is one of the most studied examples of such indirect interactions structures. Each article in Wikipedia constitutes a collective solution that results from the repeated updates of different individuals. Indirect interaction methods are usually applied to more complex problems, where the individual contributions *accumulate* over time to give rise to the collective solution. For instance, in the crisis-mapping task, a group of participants is instructed to annotate a map to indicate the zones of emergency in a particular area after a natural disaster has occurred [42]. Every group member is facing the same map, can see every modification made by the other group members, and is free to update it at any moment. Over time, the map converges to an accurate collective solution that aggregates every group member's knowledge about the situation.

Group members may also work on the collective solution *sequentially* rather than simultaneously. In a transmission chain (also called diffusion chain), each group member updates the collective solution only once before passing it to the next individual who can update it in turn, and so forth until the last individual of the group (e.g. [43]). Transmission chains have been widely used in the domain of cultural evolution [44–46][47,48], in which evolutionary anthropologists study human's ability to produce increasingly complex cultural artefacts by improving them sequentially, generation after generation. In fact, cultural evolution can be seen as a process relying, at least in part, on the principles of indirect interactions.

To illustrate these three families of aggregation methods, imagine a group of ten individuals trying to solve a jigsaw puzzle, such as the one studied by Kempe and Mesoudi [45]. The task consists in assembling together a large number of small pieces to produce a complete picture. In the 'no-interaction' case, each group member proposes an independent, possibly-incomplete solution to the problem. The individual solutions are later combined by a decision-maker to produce the collective solution. The decision-maker could, for example, follow a majority rule and fill each position of the puzzle with the piece that has been used by the majority of the group members. In the 'direct interaction' case, all ten group members would sit around the puzzle and try to solve it together by exchanging ideas and suggestions about where to put each piece and what the final picture could look like. Finally, in the 'indirect interaction' case, every group member could sequentially look at the current state of the puzzle, try to improve it by adding new pieces or moving those that are already positioned, and transmit the updated version of the puzzle to the next group member who can update it in turn. This process repeats until the tenth individual.

To date, most of the research on human collective intelligence focuses on aggregation methods that are based on direct interactions (e.g., group discussion and networked experiments), and on those where interaction is absent (e.g. the wisdom-of-the-crowd). In comparison, the mechanisms of indirect interaction methods are much less understood. Such methods are usually studied in very sophisticated forms and applied to very complex problems, which renders the basic processes of indirect interactions difficult to understand. Here, we address the question of whether indirect interactions could constitute an efficient approach for solving elementary problems. For this, we present a series of numerical simulations describing the expected performances of a transmission chain in the case of binary-choice tasks. We deliberately chose a simple aggregation method (i.e. the transmission chain, for which individuals sequentially update a collective solution) and a simple class of problems (i.e., binary-choice tasks, for which any solution is a unidimensional, binary object). We compared the performances of the transmission chain to two other aggregation methods: the majority rule and the confidence-weighted majority rule — two methods assuming the absence of interactions. By varying systematically the structure of the environment and the size of the group, we show that the best method depends on the structure of the environment. Furthermore, the analysis of two real datasets reveals which environmental structure is more likely to be found in the real world.

## Methods

In the present work, we will examine the expected performances of the transmission chain as an aggregation method, and compare it to two other methods: The majority rule, which is commonly used for aggregating judgments in discrete choice problems, and confidence-weighted majority (hereafter, weighted-majority) that extends the majority rule by taking the confidence of the individuals into account. For this, we restrain our investigations to binary-choice tasks, where only two options are possible: a correct one and a wrong one.

**Sample population.** As a starting point, we assume a large sample population in which a proportion $q_1$ of individuals would independently choose the correct solution to the problem in the absence of any interaction or social influence. Reversely, a proportion $q_0 = 1 - q_1$ of individuals in the sample population would independently choose the wrong solution. In addition, every individual's answer is associated with a confidence level $c$ describing how confident the individual is about his or her answer. We define the confidence level as a continuous value ranging from 0 to 1, where $c = 0$ refers to individuals who are very uncertain about their answer, and $c = 1$ refers to individuals who are very certain about their answer. In the simulations, we describe the confidence levels of the individuals who give the correct answer with a beta distribution $\Omega_1$ that has shape parameters $\alpha_1$ and $\beta_1$, and the confidence levels of the individuals who give a wrong answer with a beta distribution $\Omega_0$ that has shape parameters $\alpha_0$ and $\beta_0$. That is, the confidence levels follow different distributions depending on whether the associated answer is correct or wrong. For many problems, confidence can be a good proxy for accuracy, but this tendency is not systematic and often non-linear [8,10]. In fact, the shape of these two distributions depends on the nature and the statement of the problem. To begin with, we assume the two distributions $\Omega_1$ and $\Omega_0$ represented in **Fig 1**, for which correct answers are on average associated to higher confidence levels than wrong answers, but a considerable overlap exists between the two distributions (i.e. an individual with a wrong answer can possibly be more confident than an individual with the correct answer).

**Constitution of the groups.** In this environment, we compare the performances of the three aggregation methods for groups of $N$ individuals randomly selected from the sample population. Therefore, each of the $N$ individuals of one group $g$ has a probability $q_1$ to give the correct answer and a probability $q_0$ to give the wrong answer. We call $x_p$ the independent solution of one specific individual $p$ in the group, and $c_p$ the confidence level of that specific individual. We have $x_p = 1$ if the individual $p$ independently provides a correct answer, and $x_p = 0$ if that individual independently provides a wrong answer. The confidence $c_p$ of that individual is then randomly drawn from the distribution $\Omega_1$ if $x_p = 1$ (i.e. the blue distribution in **Fig 1**), and from the distribution $\Omega_2$ if $x_p = 0$ (i.e. the red distribution in **Fig 1**).

**Aggregation methods.** The outcome of each aggregation method can be computed for any given group $g$ composed of $N$ individuals. We call $M_g$, $W_g$, and $C_g$ the outcomes (0 or 1) of the majority rule, the weighted-majority rule, and the transmission chain for that particular group $g$, respectively. Furthermore, we call $M$, $W$, and $C$ the success chance of the three methods for any group $g$ of size $N$.

For a given group $g$, the outcome of the majority is $M_g = 1$ if $n_1 > n_0$ and $M_g = 0$ if $n_0 > n_1$, where $n_1$ and $n_0$ correspond to the number of group members independently choosing the correct answer (i.e. those with $x_p = 1$) and the wrong answer (i.e. those with $x_p = 0$), respectively. If $n_1 = n_0$ we choose randomly. For computing the outcome of the weighted-majority, we first measure $m_1$ corresponding to the sum of the confidence levels $c_p$ of all the group members who chose the correct answer, and $m_0$ corresponding to the sum of the confidence levels $c_p$ of all the group members who chose the wrong answer. We then have $W_g = 1$ if $m_1 > m_0$ and $W_g = 0$ if $m_0 > m_1$. If $m_1 = m_0$ we choose randomly. For the transmission chain, we first assign each group member to the chain position $p$. We call $X_p$ the current collective solution in the chain at position $p$. The first individual located at position 1 initiates the collective solution with his or her independent solution, leading to $X_1 = x_1$. For all subsequent chain positions, we assume that the individuals whose confidence level is too low are not confident enough and do not modify the collective solution. In contrary, the individuals whose confidence is sufficiently high replace the collective solution by their own independent solution if they disagree with it. Formally, we define a contribution threshold $\tau$, and have $X_p = x_p$, if $c_p \geq \tau$ (i.e. the collective solution is replaced by the individual solution if the confidence is high enough) and $X_p = X_{p-1}$, if $c_p < \tau$ (i.e. the collective solution remains unchanged if the confidence is not high enough). The outcome of the transmission chain for that particular group is then $C_g = X_N$, corresponding to the collective solution that is present in the chain after the update of the last individual.

Unlike the majority and the weighted-majority rules, the definition of the chain method requires a parameter $\tau$. This parameter represents the confidence threshold above which individuals are sufficiently confident to *contribute* to the collective solution rather than solely forwarding it to the next person. For that reason, we call $\tau$ the contribution threshold. The individual $p$ is considered to be a contributor when $c_p \geq \tau$. More specifically, we call positive contributors the group members who bring in the correct answer (i.e., those with $c_p \geq \tau$ and $x_p = 1$), and negative contributors those who bring in the wrong answer (i.e., $c_p \geq \tau$ and $x_p = 0$).

**Example.** We illustrate the outcome of each aggregation method in a simple example where $q_1 = 0.6$ and $N = 10$. That is, we constitute a group $g$ with 10 individuals randomly drawn from a sample population that contains 60% of correct answers. The confidence levels are drawn from the distributions shown in **Fig 1**. In this example, we assume a contribution threshold $\tau = 0.6$. The independent answers of each group member and their confidence are represented in **Fig 2A**. In that particular group, five individuals independently provide a correct answer, and five others provide a wrong answer. The majority rule results in a tie and has, therefore, 50% chance to yield a correct answer $M_g = 1$ and 50% chance to yield a wrong one $M_g = 0$. In that particular group, the sum of confidence levels among all individuals who provide a correct answer is $m_1 = 2.39$ and the sum of confidence levels among those who provide a wrong answer is $m_0 = 2.19$. The weighted majority rule, therefore, yields a correct answer $W_g = 1$. In the transmission chain, the first individual initialises the collective solution $X_1$ with a correct answer. At chain position 2, the individual has a wrong answer but is not confident enough ($c_p < \tau$) and thus leaves the collective solution unchanged. The first contributor with a confidence level higher than $\tau$ appears at the chain position 4. That individual replaces the collective solution by a wrong answer. The individual at position 6 is also a contributor and replaces the collective solution by the correct answer. All subsequent individuals have confidence levels lower than the contribution threshold and thus do not impact the collective solution. The correct answer set by the individual at position 6 remains unchanged until the end of the chain, leading to $C_g = 1$ for this particular example.

# Results

**Order effect.** As compared to the two majority rules, one specificity of the transmission chain is the spatial structure of the group, that is, the fact that group members are organised in a linear chain. Yet, a large number of different chains can be produced from a unique group of $N$ individuals (precisely, $N!$ different chains), depending on how the group members are ordered. What is the impact of the order in which the group members are positioned? In the example shown in **Fig 2**, only two group members are contributors (those located at chain position 4 and 6). One of them is a positive contributor whereas the other one is a negative contributor. If the positive contributor is positioned *after* the negative contributor, the chain yields a correct answer (because all subsequent individuals are neutral). Reversely, if the positive contributor is positioned *before* the negative contributor, the chain yields a wrong answer because the wrong answer overrides the correct one. In the end, this particular group of individuals has 50% chance to produce a correct answer and 50% chance to produce a wrong one, depending on the ordering of the two contributors, and irrespective of the position of the eight other group members. The ordering, therefore, has a strong impact on the expected outcome of the chain in this example. More generally, for a given group of individuals, the probability that the chain produces a correct answer is the probability that at least one positive contributor appears after the last negative contributor.

This probability equals to $N_1/(N_1+N_0)$, where $N_1$ is the number of positive contributors and $N_0$ is the number of negative contributors. The smaller the difference between $N_1$ and $N_0$ the more the outcome of the chain is sensitive to the ordering of the group members. Therefore, a given group of $N$ individuals does not always produce a unique outcome with the transmission chain. Instead, the outcome of the chain depends, to some extent, on the order in which the group members are positioned.

**Contribution threshold.** The second specificity of the chain is the presence of social influence. Unlike the two other methods, the chain does not aggregate the independant answers of every group member. Instead, it filters out the answers of those who are confident enough to change the collective solution [49]. The contribution threshold $\tau$ is thus an important parameter that determines which individuals are confident enough to contribute to the final solution. What is the impact of the contribution threshold $\tau$? To address this question, we explored the performances of the transmission chain while varying the contribution threshold from $\tau = 0$ to $\tau = 1$. For each value of $\tau$, we generated 1000 groups of size $N = 10$ with $q_1 = 0.6$ and the confidence distributions shown in **Fig 1**, and measured the success chance $C$ of the chain (i.e., how often it produced a correct answer). The result is shown in **Fig 3**. When the contribution threshold is small and approaches $\tau = 0$, every group member is confident enough to contribute. In this case, everyone overrides the previous person's solution and the outcome of the chain is simply the answer of the last individual of the chain. Because every individual has a probability $q_1 = 0.6$ of being correct, the success chance of the chain is also $C = 0.6$. Likewise, when the activity threshold approaches $\tau = 1$, none of the group members are confident enough to contribute. In this case, the answer of the first individual of the chain remains unchanged until the end of the chain. Because the first individual has a probability $q_1 = 0.6$ of providing a correct answer, success chance of the chain is also $C = 0.6$. Between these two extreme values, the performance of the chain reaches a peak for $\tau = 0.83$. At this point, the success chance of the chain is $C = 0.93$ (i.e. the chain yields a correct answer 93% of the time). In fact, the optimal threshold value maximizes the probability to pick a positive contributor, while at the same time minimizes the probability to pick a negative contributor. Note that this performance measure takes into account the order effect because groups are randomly ordered in each replication. For comparison, the success chance of the majority rule is $M = 0.73$ under these conditions, and the success chance of the weighted-majority is $W = 0.90$.

**Group size.** The weakness of the majority rule in the previous case study was the relatively small group size ($N = 10$). In fact, the majority of the individuals in the entire sample population *do* actually provide the correct answer. Yet, the majority of the $N = 10$ group members has only 73% chance to produce a correct answer because the majority *within* the group often points toward the wrong answer [25]. Hence, group size matters for the majority rule. How does it impact the outcome of the transmission chain? To address this question, we run an additional series of simulations, this time varying the contribution threshold $\tau$ as well as the group size $N$. We generate again 1000 groups of size $N$ with $q_1 = 0.6$ and the confidence distributions shown in **Fig 1**, and measure the frequency of correct answers produced by the chain for different values of $\tau$ and $N$. As **Fig 4A** shows, group size has relatively little influence on the chain performances, which increase rapidly until $N \approx 10$ and plateaus for larger group sizes. This result is consistent with the previous result showing that the ratio between positive and negative contributors is more important than the total number of contributors. In addition, the optimal contribution threshold only marginally varies between $\tau = 0.8$ and $\tau = 0.9$ with increasing group size. **Fig 4B** compares the evolution of the majority, the weighted-majority and the chain with $\tau = 0.85$ for increasing values of $N$. While the performance of the majority increases slowly with $N$, the weighted majority and the transmission chain reach higher performances for smaller group size. This weak dependency on $N$ for these two methods results from the fact that they also rely on the individuals' confidence and can thus extract the correct answer from a smaller number of individuals. The weighted-majority and the transmission chain have relatively similar performances in this environment, with the chain converging slightly faster to its best performance, and the weighted majority converging slower but reaching a slightly higher performance level.

**Structure of the environment.** In order to generalize our findings, we explored the performances of the three methods for different proportions of correct answers $q_1$ in the sample population, and different confidence distributions. It is difficult, however, to vary systematically the confidence distributions, because these distributions depend on a total of four parameters (i.e., the shape parameters $\alpha_1$, $\beta_1$ and $\alpha_0$, $\beta_0$). Therefore, we computed a single confidence indicator from the two distributions $\Omega_1$ and $\Omega_0$, that we called the *confidence utility u*. The confidence utility measures the probability that an individual with the correct answer has a higher confidence level than an individual with the wrong answer. In other words, $u$ is the probability that a value drawn from $\Omega_1$ (the confidence associated to correct answers) is higher than another value drawn from $\Omega_0$ (the confidence associated to wrong answers). For the simulations, we systematically varied the mean values of $\Omega_1$ and $\Omega_0$ from 0.1 to 0.9 and computed each time the corresponding confidence utility $u$. In addition, we also varied the proportion of correct answers $q_1$ from 0 to 1. For each combination of $u$ and $q_1$, we generated 1000 groups of size $N = 10$ and measured the performance of the three aggregation methods. For the sake of simplicity, we chose the contribution threshold $\tau$ that maximises the success chance of the chain for each combination of $u$ and $q_1$, such that our results represent the best-case scenario for the chain. The results are shown in **Fig 5**. Clearly, the majority rule does not depend on $u$ and has success chances that approach $M = 1$ when $q_1 > 0.5$ and $M = 0$ when $q_1 < 0.5$. The majority rule amplifies the dominant view in the population. The weighted-majority has a similar pattern but also relies on the group members' confidence. Therefore, the weighted-majority can yield a correct answer even when the population accuracy $q_1$ is slightly lower than 0.5, but only when the confidence is a useful cue (i.e., when $u$ approaches 1). Reversely, if the confidence is misleading (i.e. when $u$ approaches 0), even a population accuracy higher than 0.5 is not always sufficient to yield a correct collective answer. The performance of the transmission chain exhibits a similar tendency: When confidence is a useful cue ($u > 0.5$), the chain performs well, even for extremely low values of $q_1$. When confidence is less reliable ($u < 0.5$), the chain performance equals the population accuracy. This is due to the fact that, when confidence is misleading, the best contribution threshold is such that nobody contributes to the collective solution (i.e. very high value of $\tau$), and the final solution is simply the first individual's solution. Hence the chain performs relatively well where the majority performs very good, and relatively bad where the majority performs very bad. The best aggregation method, therefore, depends on the values of $q_1$ and $u$. Roughly speaking, the majority is a better method in the upper-left quarter of the parameter spaces ($q_1 > 0.5$ and $u < 0.5$), the weighted-majority is better in the upper-right quarter ($q_1 > 0.5$ and $u > 0.5$), and the chain is better in the lower-right quarter ($q_1 < 0.5$ and $u > 0.5$). The lower-left quarter ($q_1 < 0.5$ and $u < 0.5$) is the most difficult environment because the population accuracy and the confidence both point to the wrong answer. In this case, the chain is the least bad method because it preserves a small chance of success where the majority and the weighted-majority tend to be systematically wrong.

**Empirical data.** An important question that arises from these simulations is what do real environments look like? Knowing the values of $q_1$ and $u$ for a given class of problems would help finding out what is the most efficient aggregation method to use. To address this question, we reanalysed two datasets of previously published experiments in which people were facing a series of binary-choice tasks. In the first dataset [50], 109 participants were instructed to indicate which of two cities has a larger population, across 1000 pairs of cities. For each pair of cities, the participants indicated their answer and their confidence level on a continuous scale between 0.5 and 1. In the second dataset [22,51], a total of 40 physicians evaluated 108 cases of skin lesions and were instructed to evaluate whether the lesion is cancerous or not-cancerous, and to indicate their confidence level on a Likert scale from 1 to 4. For each of the 1000 instances of the cities dataset and for each of the 108 instances of the doctors dataset, we estimated the population accuracy $q_1$ by measuring the proportion of correct answers among the respondents, and the confidence utility $u$ by measuring the probability that an individual who gave a correct answer reported a higher confidence than an individual who gave a wrong answer. For this, we randomly sampled (1000 times, with replacement) one individual among those who gave a correct answer and one among those who gave a wrong answer, and looked at the frequency at which the first has a higher confidence than the second. In addition, we computed the success chance of the three aggregation method, with groups of size $N = 10$. The results are presented in **Fig 6**. The first interesting element is the striking diversity of environmental structures. Within each domain, the observed values of $u$ and $q_1$ vary almost uniformly between 0 and 1. Nevertheless, a correlation is visible between these two variables: confidence is a relevant cue when most people give the correct answer, and a misleading cue when most people give the wrong answer. In other words, confidence tends to be indicative of consensuality rather than accuracy – an important result that has been already suggested in previous research [8,52]. Consequently, confidence adds little information to what is already available by just looking at the individuals' answers. Most cases, therefore, lie in the upper-right and lower-left quarter of the parameter space. As expected from our previous simulations, the majority and weighted-majority perform best in the upper-right quarter, because both the number of individuals and the confidence levels indicate the correct answer. The chain also performs relatively well in this area of the parameter space (see **Fig 5** and **Figs S1 and S2** in the supplementary material), but not as good as the two majority rules. In the lower-left quarter, however, all methods exhibit poor performances, because all available information is misleading. In this area, the chain does not amplify the misleading information as the two majority rules do and, thus, preserves a small chance to yield the correct answer.

When using a single method for all the instances of a task, very little difference of performance exist between the three methods. The strengths and weaknesses of each method compensate each other at the aggregate scale, leading to similar overall performances (For the "cities dataset": M=74%, W=75%, and C=75%; for the "doctors dataset": M=65%, W=65%, and C=67%). Ideally, one would adaptively choose the most efficient aggregation method for each new instance of the task, depending on the expected values of $q_1$ and $u$ for that instance, but the diversity of observed structures makes it difficult, if not impossible, to anticipate the nature of an upcoming instance of the problem.

## Discussion

We have defined and studied the performance of a new aggregation method – the transmission chain – as a tool to improve collective decision-making in binary choice tasks. The transmission chain relies on processes of indirect interactions, for which group members sequentially try to improve a common collective solution without directly interacting with one another. The chain exploits the *default* heuristic – a common bias in people's decision-making: When given a decision to make and a default option, people tend to choose the default option unless they have good reasons not to do so [53,54]. In the chain, the "default option" is the current state of the collective solution that has been produced by the preceding group members. Thus, most people choose the default option unless they strongly disagree with it. In the end, the chain functions as a confidence-based filter: the group members who are the most confident contribute to the collective solution whereas those who are unsure have no impact on the collective outcome. When comparing the transmission chain to the majority and the weighted-majority rules, we have found that the chain only outperforms the two other methods for problems of a specific nature, namely, when confidence is a reliable indicator *and* when the most people give a wrong answer. In the two datasets that we have analysed, however, instances of such nature are rare.

It is important to note that many behavioural components of our simulations remain uncertain. For instance, we have always studied the best-case scenario in our simulations, where the contribution threshold $\tau$ is set to the optimal value for each instance of the problem. However, the contribution threshold is not a parameter but a behavioural variable, for which the experimenter has little control. The exact value of the threshold and whether it is well calibrate is an issue that needs to be addressed experimentally. In **Figs** S1 and S2, we compared the chain performances when $\tau$ is adjusted for every instance of the problem and when $\tau$ is fixed for all instances of the problem. Overall, the difference is minor.

A multitude of implementations can be imagined for the transmission chain. Group members can act simultaneously rather than sequentially (i.e., all group members can see the current state of the collective solution at the same time and update it at any moment), which should mitigate the order effect. Beside, the collective solution can take various forms. Instead of considering only the solution of the predecessor, group members could be exposed to the $k$ previous solutions, or an aggregated form of the $k$ previous solutions. This variation should create a "memory" in the chain. This memory could allow the collective solution to resist to the detrimental influence of a few outliers, but at the same time, increase the risk of opinion herding and groupthink [5,37,55]. Finally, participants could be informed about their position in the chain and about the number of other group members who have previously contributed to the collective solution. This would strengthen the weight of the collective solution at the end of the chain, equivalent to gradually increasing the contribution threshold with chain position in simulations.

With regard to the existing literature and to the present results, the transmission chain seems to be more suited to cumulative, multidimensional problems than simple binary choice tasks. In fact, in complex tasks like writing a Wikipedia article, the group members' contributions can take various forms and apply to many different components of the collective solution. The contributions to a Wikipedia article, for example, can range from correcting a minor typographical error to changing the entire structure of the article or adding new content to an existing article. As such, the contributions of the group members accumulate over time but rarely override each other (except, e.g. in the case of "edit wars", see, [56]). The same applies to problems addressed in cultural evolution [46]. In contrast, binary choice tasks are limited to a single possible action of the group members: switch the collective solution to the other answer or not. Thus, every new contribution necessarily overrides the previous ones. The cumulative property is therefore absent. Furthermore, the uncertainty surrounding the contribution threshold is important. If the contribution threshold is well calibrated, the chain is a good candidate for certain types of binary choice tasks. Otherwise, the chain will arguably not outperform a simple majority. The next step, therefore, is to examine the value of the threshold experimentally, explore more sophisticated implementations of the design, and extend the methodology to more complex problems.

# Acknowledgements

We thank Stefan Herzog, Ralf Kurvers, and Max Wolf for fruitful discussions. We are grateful to Shuli Yu, Tim Pleskac, Ralf Kurvers and Guiseppe Argenziano for sharing their experimental data. This research was supported by a grant from the German Research Foundation (DFG) as part of the priority program on *New Frameworks of Rationality* (SPP 1516) awarded to Ralph Hertwig and Thorsten Pachur (HE 2768/7-2). The funders had no

# Figures

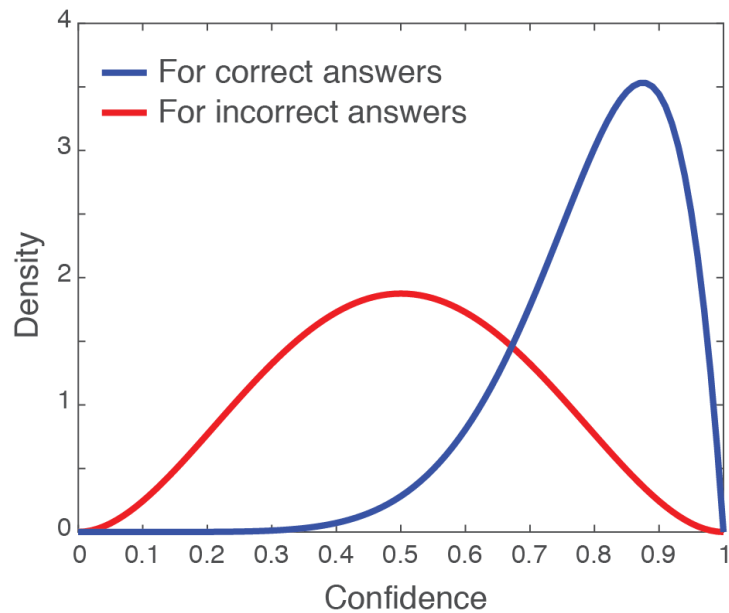

**Figure 1**: **Description of the environment**. Assumed distributions of confidence among the individuals who provide the correct answer to the problem (in blue), and among those who provide a wrong answer to the problem (in red). The interval of confidence values ranges from $c = 0$ (very uncertain) to $c = 1$ (very certain). The blue and red distributions are beta distributions with shape parameters $\alpha_1 = 8$ and $\beta_1 = 2$ (mean value: 0.8), and $\alpha_0 = 3$ and $\beta_0 = 3$ (mean value: 0.5), respectively. In the simulations, a proportion $q_1$ of the sample population gives the correct answer and have confidence levels drawn from the blue distribution, and a proportion $q_0 = 1 - q_1$ of the sample population gives a wrong answer and have confidence levels drawn from the red distribution. In empirical data, the shape parameters of the blue and red distributions depend on the nature of the task.

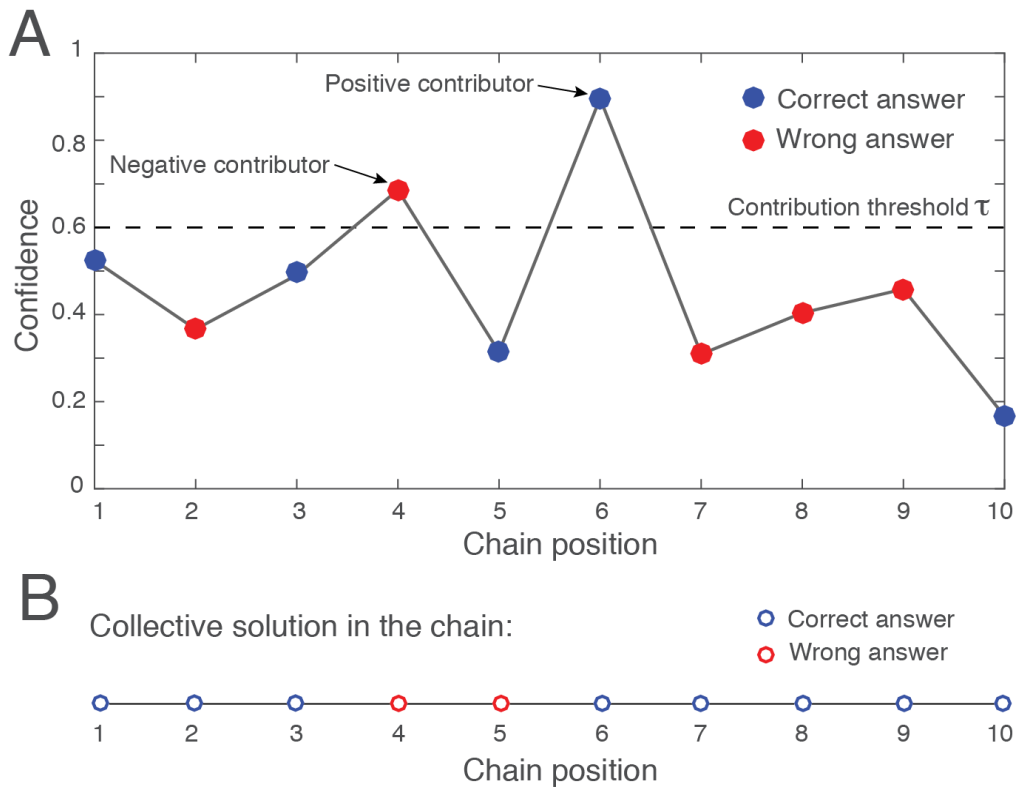

**Figure 2**: **Illustration of a transmission chain.** For this case study, we assume a group size of $N = 10$, and a proportion of correct answers $q_1 = 0.6$. (A) The $N$ individuals are randomly drawn from the sample population and assigned to a random position in the chain (the red and blue dots). Among them, five individuals have the correct answer (i.e., the blue dots at chain positions 1, 3, 5, 6, and 10), and five individuals have a wrong answer (i.e., the red dots at chain position 2, 4, 7, 8, and 9). The black dashed line represents the activity threshold $\tau = 0.6$ indicating the confidence level above which individuals contribute to the collective solution. The "contributors" replace the current collective solution by their own solution. Individuals who do not contribute (i.e., those with a confidence level lower than $\tau$) leave the collective solution unchanged. (B) The resulting collective solution in the chain at each position. The blue open circles indicate a correct answer, and the red open circles indicate a wrong one. In this example, the individual at chain position 1 initialises the collective solution with a correct answer. The collective solution remains unchanged until the contributor at chain position 4 replaces it by a wrong answer. The individual at position 6 is also a contributor and restores the correct answer. All other individuals have no impact on the collective solution. In this example, the chain generates a correct solution.

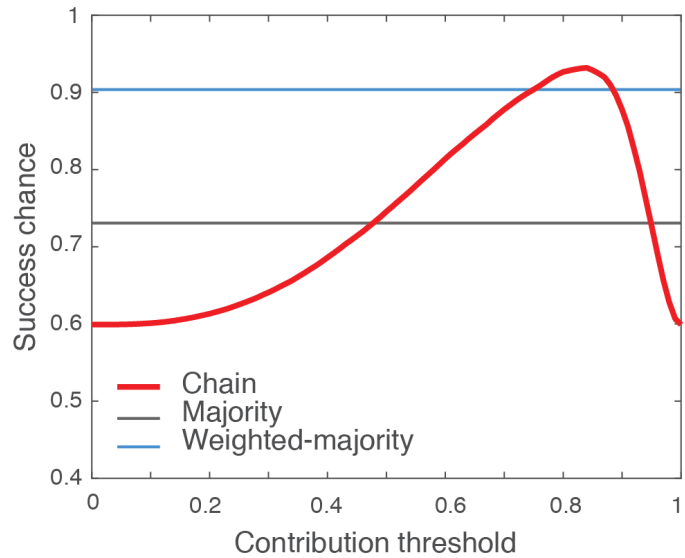

**Figure 3**: **Impact of the contribution threshold $\tau$.** The red line indicates the probability that the chain generates a correct solution for different values of the contribution threshold $\tau$, assuming a group size of $N = 10$, a proportion of correct answers $q_1 = 0.6$ and the confidence distributions shown in **Fig 1**. In these conditions, the optimal value for the contribution threshold is $\tau = 0.83$, for which the chain produces the correct solution 93% of the time. The grey and blue lines indicate the success chances of the majority and the weighted-majority rules, respectively.

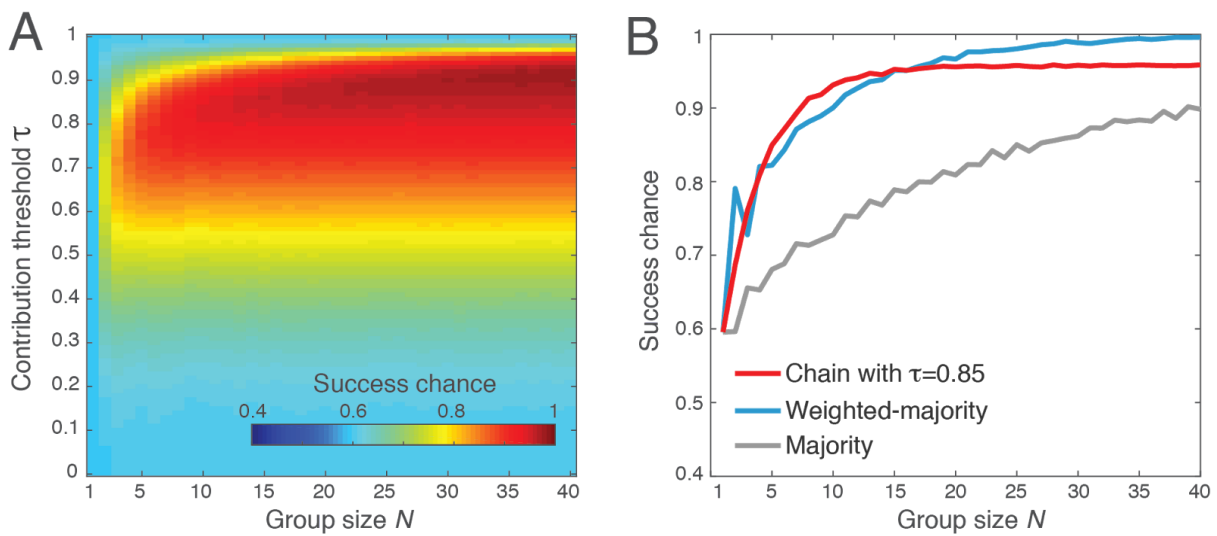

**Figure 4**: **Impact of the group size $N$.** (A) The color-coding indicates the probability of success of the chain method as a function of the group size $N$ and the contribution threshold $\tau$. The column of values at $N = 10$ corresponds to the red curve shown in **Fig 3**. (B) Comparison of the performance of the chain method with the contribution threshold set to $\tau = 0.85$ (in red), the majority rule (in grey), and the weighted-majority rule (in blue). These results are computed assuming a proportion of correct answer $q_1 = 0.6$ and the confidence distributions shown in **Fig 1**.

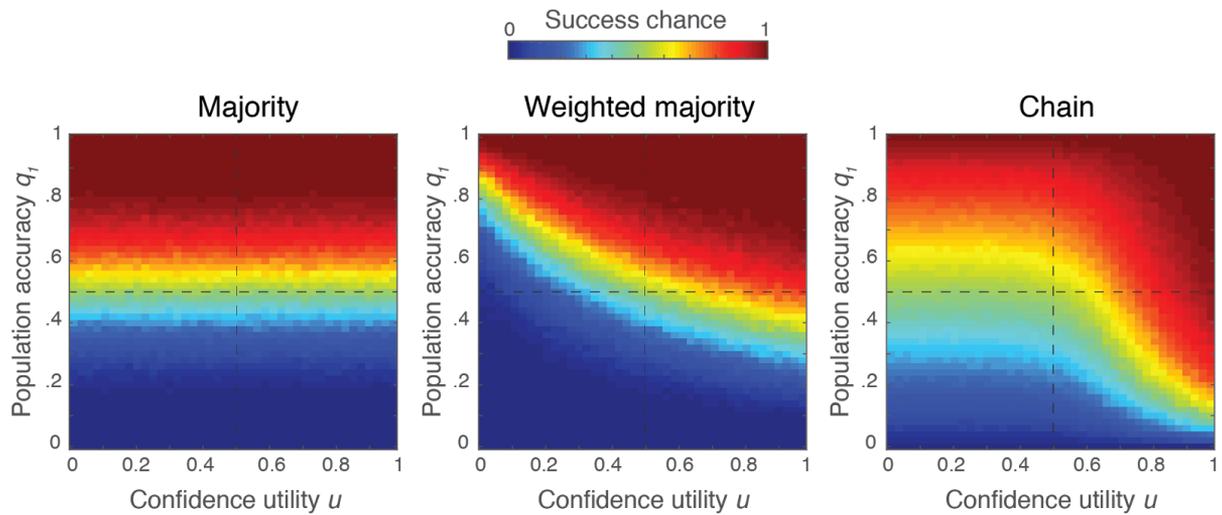

**Figure 5: Impact of population structure**. Expected performance of the majority, the weighted-majority, and the transmission chain as a function of the proportion of correct answers $q_1$ in the sample population and the confidence utility $u$. The confidence utility is the probability that an individual with a correct answer reports a higher confidence level than another individual with a wrong answer. The group size is $N = 10$.

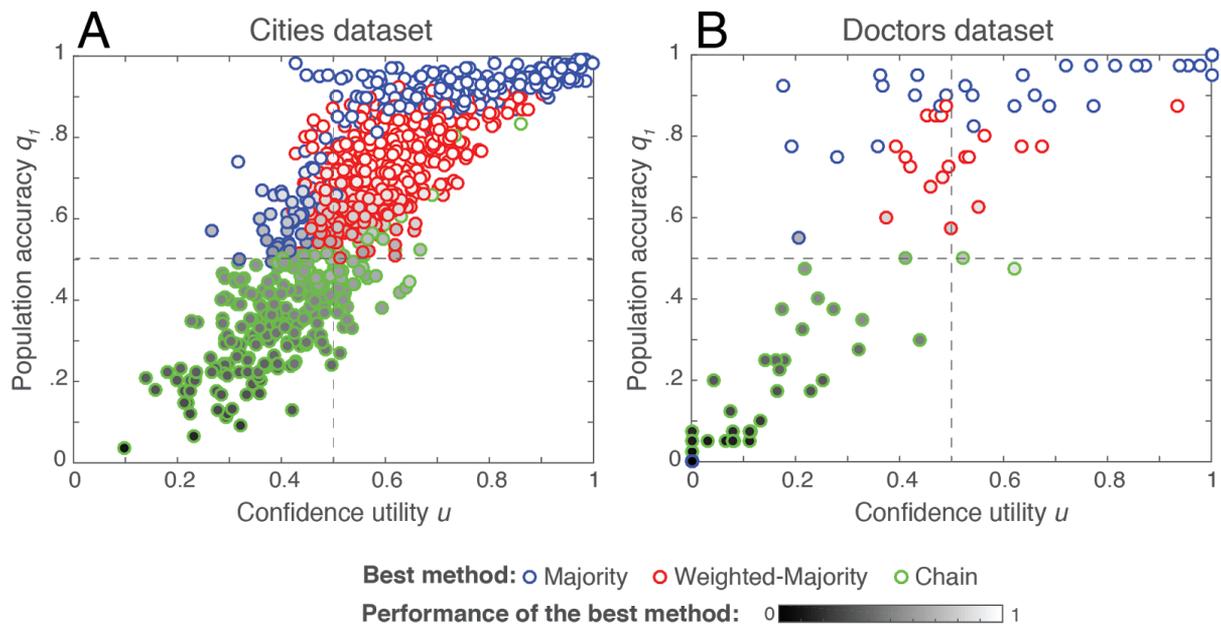

**Figure 6**: **Population structure and corresponding best method for two binary-choice task studies.** (A) Experimental participants evaluating which of two cities has a larger population. The task is repeated across 1000 different pairs of cities. Each point in the graph corresponds to one instance of the task (i.e., one pair of cities). (B) Dermoscopists evaluating 108 cases of skin lesions and evaluating whether the lesion is cancerous or not-cancerous. Each point in the graph corresponds to one medical case. In (A) and (B) the position of each point indicates the proportion of respondents (i.e., participants or doctors) who provided the correct answer ($y$-axis), and the confidence utility ($x$-axis) for that case. The border colour of each point indicates the aggregation method that performs best in this particular case (the majority in blue, the weighted-majority in red, and the chain in green). Cases for which several methods perform equally good are represented in blue if the majority rule is one of them and in red if the weighted-majority is one of them. In addition, the grey-scale colour inside each point indicates the success chance of the best performing method, ranging from 0 (in black) to 1 (in white). These results are calculated assuming 1000 groups of $N = 10$ individuals randomly sampled from the pool of available respondents.

# Supplementary Material

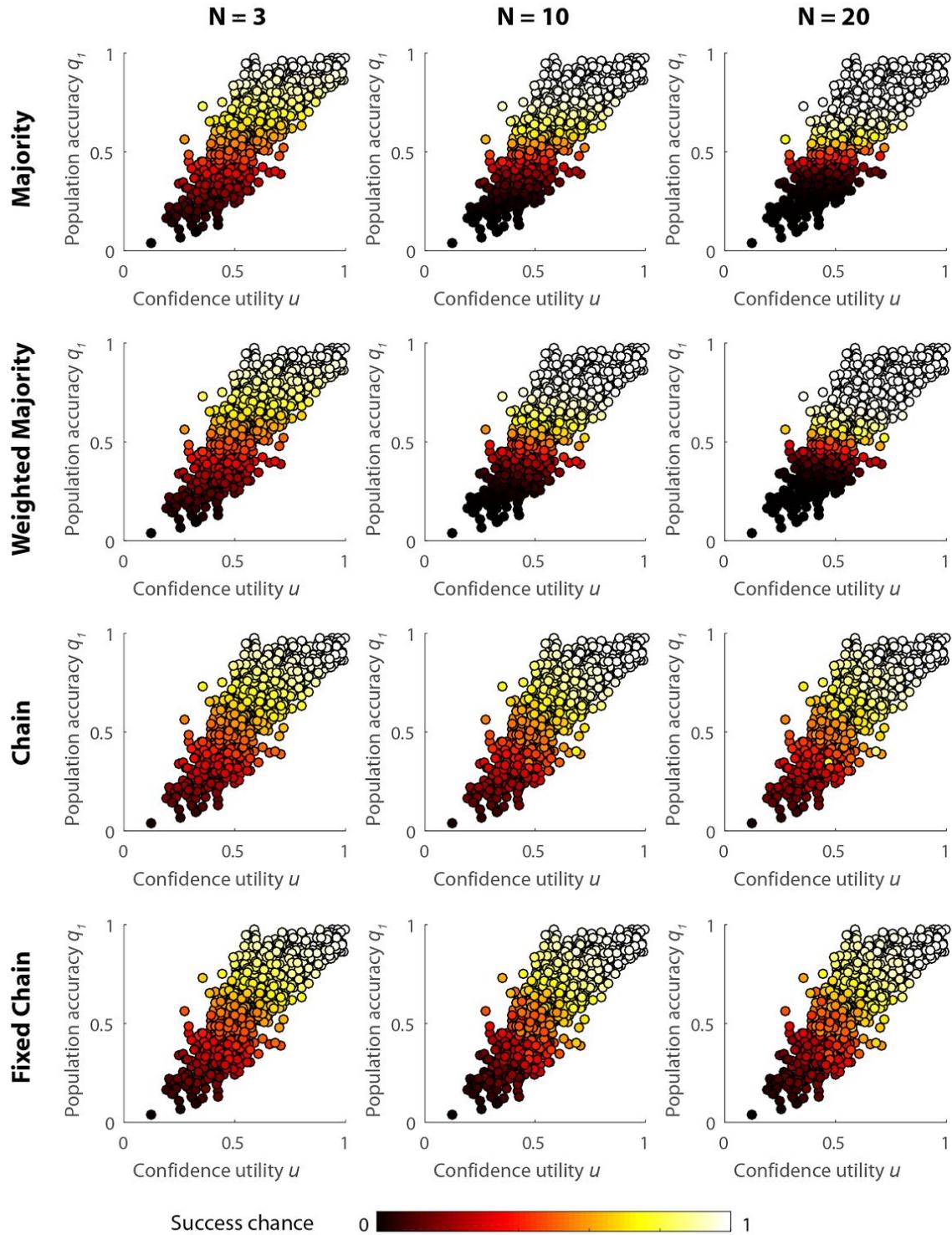

**Figure S1:** Performance of different aggregation method for groups of size $N = 3, 10, 20$ in the cities dataset. Each line corresponds to one aggregation method. The fixed chain has the contribution threshold fixed to $\tau = 0.8$ for all instances of the task (in contrast to the chain for which the contribution threshold is adjusted between instances of the task). Each point in the graphs corresponds to one city comparison task. The position of each point indicates the proportion of participants who provided the correct answer (*y*-axis), and the confidence utility (*x*-axis) for that task . The colour of each point indicates the success chance of each method for that particular task.

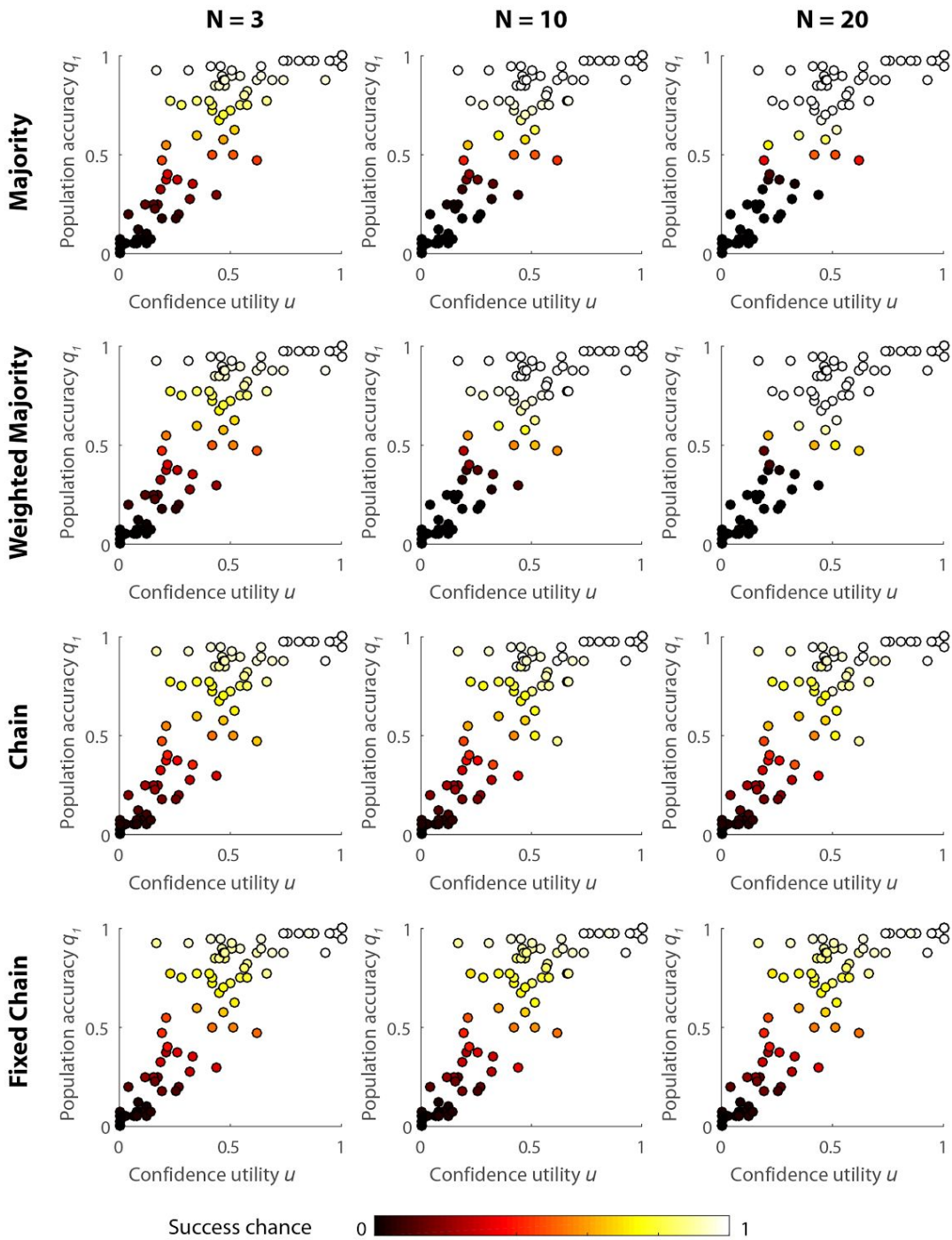

**Figure S2:** Performance of different aggregation methods for groups of size $N = 3, 10, 20$ in the doctors dataset. Each line corresponds to one aggregation method. The fixed chain has the contribution threshold fixed to $\tau = 3$ for all instances of the task (in contrast to the chain method for which the contribution threshold is adjusted between instances of the task). Each point in the graphs corresponds to one medical case. The position of each point indicates the proportion of doctors who provided the correct answer (*y*-axis), and the confidence utility (*x*-axis) for that case . The colour of each point indicates the success chance of each method for that particular case.